\documentclass{ifacconf}                           
\usepackage{natbib}        

\usepackage{flushend}
\usepackage{lscape}
\usepackage{bigints}
\usepackage{graphicx} 
\usepackage{psfrag}
\usepackage{amsfonts}
\usepackage{amssymb}
\usepackage{amsmath}
\usepackage{graphicx}
\usepackage{psfrag}
\usepackage{ae,epsfig}
\usepackage{enumerate} 

\usepackage{epstopdf}
\usepackage{tikz}
\usetikzlibrary{shapes,arrows}
\usetikzlibrary{positioning,fit}
\usepackage{color}
\usetikzlibrary{positioning,fit}
\usepackage{bm}
\usetikzlibrary{calc}
\usetikzlibrary{matrix}

\newcounter{defcount}

\newtheorem{definition}[defcount]{Definition}

\begin{document}
\begin{frontmatter}

\title{Structural Identifiability of a Pseudo-2D Li-ion Battery Electrochemical Model\thanksref{footnoteinfo}} 

\thanks[footnoteinfo]{Work supported by the Engineering and Physical Sciences Research Council through grant EP/P005411/1- ``Structured electrodes for improved energy storage.''}

\author[First]{Ross Drummond} 
\author[First]{Stephen R. Duncan} 

\address[First]{Department of Engineering Science, University of Oxford, 17 Parks Road, OX1 3PJ, Oxford, United Kingdom. \\Email: \{ross.drummond, stephen.duncan@eng\}@eng.ox.ac.uk.}

\begin{abstract}                
 Growing demand for fast charging and optimised battery designs is fuelling significant interest in electrochemical models of Li-ion batteries. However, estimating parameter values for these models remains a major challenge. In this paper, a structural identifiability analysis was applied to a pseudo-2D Li-ion electrochemical battery model that can be considered as a linearised and decoupled form of the benchmark Doyle-Fuller-Newman model. From an inspection of the impedance function, it was shown that this model is uniquely parametrised by 21 parameters, being combinations of the electrochemical parameters like the conductivities and diffusion coefficients. The well-posedness of the parameter estimation problem with these parameters was then established. This result could lead to more realistic predictions about the internal state of the battery by identifying the parameter set that can be uniquely identified from the data.

\end{abstract}


\end{frontmatter}

\section*{Introduction}

As the importance of electrical energy storage continues to grow and Li-ion batteries numbers keep on increasing, this technology is becoming ever more mature. The sheer number of Li-ion batteries being produced is leading to significant price reductions (\cite{curry}), due primarily to the economics of scale, meaning that any new battery chemistry hoping to be commercialised 
will have to overcome increasingly steep economic barriers before entering the mass-market. As such, at least in the near future, the next wave of battery innovation is likely to come from optimising existing technology, rather than from the introduction of radically new chemistries. Control engineering will play a pivotal role in this optimisation.

Electrochemical models are important tools for optimising battery use, both for control purposes within a battery management system (\cite{chaturvedi}) and for design. These models provide a rich description of the battery's response in terms of snapshots of its internal electrochemical state; however, their accuracy critically relies upon the electrochemical parameter values. Obtaining accurate estimates of these parameters has thus emerged as a crucial research topic in recent years. Unfortunately, the relative complexity of electrochemical models makes estimating their parameter values difficult. This difficulty was illustrated in \cite{moura_genetic} where it took approximately three weeks’ worth of computation on five quad-core Intel Q8200 computers to estimate the parameters of the benchmark Doyle-Fuller-Newman (DFN) model (\cite{Newman1}). To bring some clarity to this problem, \cite{adrien_ident} recently approached the parameter estimation problem from a different perspective by performing a structural identifiability analysis of the simplified single particle model (SPM), determining the unique six parameters fully describing the SPM's response. In such a way, the well-posedness of the parameter estimation problem for this simplified model could be established. This result built upon existing literature in this area, including \cite{pozzi} where sensitivity functions for a polynomial approximations of the single particle model with electrolyte (SPMe) were obtained, \cite{park} which considered experiment design for ranking the sensitivity of each electrochemical parameter, \cite{xinfan} where analytic sensitivity functions for the SPM's parameters were derived from Pad\`{e} approximations of the impedance and \cite{randles} that analysed the structural identifiability of Randles circuit battery models. 

Motivated by \cite{adrien_ident}, this paper develops a framework towards a structural identifiability analysis of the DFN electrochemical Li-ion battery model (\cite{Newman1}). The DFN model is widely considered as a benchmark micro-scale model for Li-ion batteries from which simplifications like the SPM (\cite{SPM1,SPM2}) and the SPMe (\cite{SPMe,marquis,giles}) are derived. But, estimating its parameter values in a methodical way is challenging (\cite{moura_genetic}). This paper details preliminary results on this problem by performing a structural identifiability analysis of a decoupled and linearised form of the DFN model which can be considered as the SPMe with added double-layer effects. 

To perform the structural identifiability analysis, the DFN model was first simplified into a linearised form with dynamics decomposed into three elements: solid-state diffusion, bulk electrolyte mobility and the charge transfer resistance caused by the relaxation of the overpotential. This simplification enables a tractable analysis, with the generalised impedance functions developed in \cite{sikha} being too complex, right now, to determine structural identifiability. It is shown that this decoupled DFN model is uniquely parametrised from current/voltage data by 21 parameters formed from combinations of the electrochemical parameters (see Table \ref{tab:par}). It is hoped that the results of this work will provide the theoretical underpinning behind a generalised parameter estimation method of the DFN model that will not rely upon substantial a prior knowledge of the cell's makeup. The need for accurate, and recursive, estimates of the DFN model's parameters is expected to grow still further as the importance of fast charging (\cite{fast}) and battery design become ever clearer. For these applications, the SPM provides neither sufficient accuracy nor richness. 

 \section*{Notation}
Definitions of the electrochemical model states and parameters are given in Table \ref{tab:nom}. Special attention is paid to the definition of the spatial domains, identified by the subscript $k \in \{1, 2, 3\}$, $k = 1$ denoting the anode, $k = 2$ the separator and $k = 3$ the cathode. For instance, the point $x = L_{12}$ defines the anode/separator boundary and $x = L_{23}$ is the separator/cathode boundary.  $\Omega_1~:=~\{x:~0~<~x~<~L_{12}~\}$ denotes the spatial domain of the anode, $\Omega_2~:=~\{x:~L_{12}~<~x~<~L_{23}~\}$ that of the separator and $\Omega_3~=:~\{x:~L_{23}~<~x~<~L~\}$ that of the cathode.  When a parameter takes different values in each domain then the subscript $k$ is added, for instance $\varepsilon_1$ is the porosity in the anode. When these parameter subscripts are missing, like in \eqref{ss_dyns}, this indicates that the equation holds in all domains. This notation is adopted for compactness to avoid duplication of the equations as much as possible.  
 
 For spatio-temporal variables (e.g. $c_e(x,t)$), tilde will denote variations (e.g. $\tilde{c}_e(x,t))$ around an equilibrium denoted by an $^*$ (e.g. $c^*_e$). In particular, the expansions
 \begin{subequations}\begin{align}
 c_e(x,t)  & = c_e^* + \tilde{c}_e(x,t), \\
  u_s(x,r,t)  & = \tilde{u}_s(x,r,t) + \frac{r}{R_s}\tilde{u}_s^{\text{surf}}(x,t),\\
   u_s(x,r,t)  & = rc_s(x,r,t) 
 \end{align}\end{subequations}
are used. The surface concentrations are $ c_s^{\text{surf}}(x,t) = c_s(x,R_s,t) $ and the transformed surface concentrations are $ u_s^{\text{surf}}(x,t) = u_s(x,R_s,t) $. Overbars denote Laplace transforms of signals, e.g. $\bar{\tilde{c}}_e(x,s)$.

\section{Problem formulation}

Structural identifiability of a decomposed form of the DFN electrochemical model for Li-ion batteries is considered. In each electrode, the DFN model electrochemical equations  are
\begin{subequations}\label{DFN_eqns_orig}\begin{align}
\frac{\partial u_s(x,r,t)}{\partial t} &= D_s \frac{\partial ^2u_s(x,r,t)}{\partial r^2}, \\
 \frac{\partial c_\text{e}(x,t)}{\partial t}  & =  D_\text{e}\frac{\partial^2c_e(x,t) }{\partial^2 x}  + \frac{a_\text{s} \left(1-t_+\right)}{\varepsilon} j(x,t), \\
\eta(x,t) &  = \phi_\text{s}(x,t) - \phi_\text{e}(x,t) - U (u_\text{s}^\text{surf}(x,t)), \label{eta}\\
i_{0}   =   k  \mathcal{F}     (c_\text{s}^\text{max} &    -   c_\text{s}^\text{surf}(x,t) )^{\alpha_\text{a}}  c_\text{s}^\text{surf}(x,t) ^{\alpha_\text{c}} c_\text{e}(x,t) ^{\alpha_\text{a}} , \\
j(x,t) & = \frac{i_0(x,t)}{\mathcal{F}} 
\sinh \left( \frac{\mathcal{F}}{RT} \eta(x,t)\right) ,\label{j_bv} \\
\frac{i_\text{s}(x,t)}{\sigma^{\text{eff}}} &  = - \frac{\partial \phi_\text{s}(x,t)}{\partial x}, \\
\frac{i_\text{e}(x,t)}{\kappa} &  = 
- \frac{\partial \phi_\text{e}(x,t)}{\partial x} 
+ K \frac{\partial  \tilde{c}_\text{e}(x,t)}{\partial x}, \label{ie1} \\
 i(t) & = i_\text{s}(x,t) + i_\text{e}(x,t) , \\
\frac{\partial i_\text{e}(x,t)}{\partial x}  & = a_\text{s} \mathcal{F} j(x,t),\label{div_eqn}\end{align}\end{subequations}
and, in the separator, the governing equations are simply
\begin{subequations}\begin{align}
 \frac{\partial c_\text{e}(x,t)}{\partial t}  & =   D_\text{e}\frac{\partial^2 c_e(x,t)}{\partial x^2}, \\
\frac{i(t)}{\kappa} &  = 
- \frac{\partial \phi_\text{e}(x,t)}{\partial x} 
+ K \frac{\partial \tilde{c}_\text{e}(x,t)}{\partial x}. \label{ie2}
\end{align}\end{subequations}
In \eqref{ie1}, \eqref{ie2}, it was assumed that $c_e^* \gg 0$ so that MacInnes' equation could be approximated with  $\frac{\partial \ln(c_e)}{\partial x} \approx \frac{1}{c_e^*}\frac{\partial \tilde{c}_e}{\partial x} $ since a linearised model is eventually  used for the impedance response. This assumption holds for most concentrated electrolytes. To obtain statements on structural identifiability of this electrochemical model, three further assumptions are applied to the DFN model equations.

\textbf{Assumption 1}: \textit{All parameters are time independent}. Time independence allows the analysis to be conducted in the frequency domain and the statements on structural identity to be derived, although, in doing so, thermal and other state-dependent effects are ignored. 

\textbf{Assumption 2}: \textit{The applied currents are small enough so that a small signal analysis can be performed}. The states then do not deviate significantly from their equilibria $c_e(x,t) \approx c_e^*$, $c_s^{\text{surf}} \approx c_s^{\text{surf},*}$ and $\partial \eta \approx \partial \phi_{dl}$. 

\textbf{Assumption 3}: \textit{Simplified relations for the Butler-Volmer inter-facial reaction rate \eqref{j_bv} are used}. As adopted in \cite{ong} for the analysis of the fast overpotential dynamics \eqref{eta_dyns}, $j(x,t)$ is simplified through a linearisation of \eqref{j_bv} with small deviations in concentrations from equilibrium assumed $i_0 \approx i_0^*$ (with $i_0^*$ denoting the ion exchange current evaluated at $c_e = c_e^*$ and $c_s^{\text{surf}} =  c_s^{\text{surf},*}$) so
 \begin{align}\label{j_lin}
j \approx \frac{i_0^*a_sF}{RT}\eta(x,t).
\end{align} 
For the slower particle \eqref{dyns_part} and bulk electrolyte \eqref{dyns_ce} dynamics, constant reactivity across the electrode is assumed 
\begin{align}\label{j_const}
j^*(t)  & = \frac{-I(t)}{a_sL\mathcal{F}A}, ~x \in \Omega_1, ~j^*(t)  = \frac{I(t)}{a_sL\mathcal{F}A}, ~ x \in \Omega_3.
\end{align}
Constant reactivity across the electrode is used in the control-orientated SPM and SPMe  models (\cite{SPMe,marquis}).

 \begin{table}
\centering 
\renewcommand{\arraystretch}{1.3} 
\begin{tabular}{c|l | l}
\hline 
\multicolumn{3}{c}{Spatial Variables} \\
\hline
$x$ & Battery spatial variable. & m\\
$r$ & Particle spatial variable. & m\\
\hline 
\multicolumn{3}{c}{Variables} \\
\hline
$c_s(x,r,t)$ & Particle Li concentration. & mol m$^{-3}$. \\
$c_s^{\text{surf}}(x,t)$ & Li surface concentration. & mol m$^{-3}$. \\
$ c_e(x,t)$ & Effective electrolyte conc. & mol m$^{-3}$. \\
$u_s(x,r,t)$ & Transformed concentration.  & mol m$^{-2}$. \\
$u_s^{\text{surf}}(x,t)$ & Transformed surface conc. &  mol m$^{-2}$. \\
$\phi_s(x,t)$ & Potential in the solid. & V.\\
$\phi_e(x,t)$ & Potential in the electrolyte. & V.\\
$\phi_{dl}(x,t) $ &$ \phi_s(x,t)-\phi_e(x,t)$. & V.\\
$i_s(x,t)$ & Current in the solid. & A m$^{-2}$\\
$ i_e(x,t)$ & Current in the electrolyte. & A m$^{-2}$. \\
$j(x,t)$ & Reaction rate. & A m$^{-3}$. \\
$\eta(x,t)$ & Overpotential. & V. \\
$U(u_s^{\text{surf}})$ & Open circuit potential. & V. \\
$ i(t)$ & Applied current density.& A m$^{-2}$. \\
$V(t)$ & Measured Voltage & V. \\
\hline 
\multicolumn{2}{c}{Parameters} \\
\hline
$c_s^{\text{max}}$ & Max. Li particle concentration. & mol m$^{-3}$. \\
$D_s$ & Particle diffusion coefficient. & m$^2$ s$^{-1}$.\\
$R_s$ & Particle radii. &  m. \\
$L_{{k}}$ & Length of domain $k \in \{1,\,2,\,3 \}$. & m. \\
$L_{{k}}$ & Separator boundary $k \in \{12,\,23 \}$. & m. \\
$L$ & Total length of the battery. & m. \\
$\varepsilon$ & Porosity coefficient. &  \\
$D_{\text{e}}$ & Effective diffusion coefficient. & m$^2$ s$^{-1}$. \\
$a_s$ & Specific interfacial area. & m$^{-1}$. \\
$t_+$ & Transference number. &  \\
$\mathcal{F}$ & Faraday's constant. & C mol$^{-1}$. \\
$\alpha_{c}$/$\alpha_c$ & Anode/Cathode transfer coefs. & \\
$R$ & Universal gas Constant. &  mol$^{-1}$ K$^{-1}$. \\
$T$ & Temperature.  & K.\\
$k$ & Exchange rate parameter. & m$^{2.5}$mol$^{-0.5}$s$^{-1}$. \\
$\sigma$ & Electronic conductivity. & S m$^{-1}$. \\
$\kappa$ & Ionic conductivity. &  S m$^{-1}$. \\
$R_{ctc}$ & Contact resistance.  & $\Omega$. \\
$K$ & $\frac{2 \left(1 - t_+ \right) RT}{c_e^*\mathcal{F}} $ & \\
$A_{\text{cc}}$ & Current collector surface areas. & \\
\hline 
\multicolumn{3}{c}{Sub and super scripts} \\
\hline
1 & Denotes anode parameters.  & \\
2 & Denotes separator parameters. & \\
3& Denotes cathode parameters. &  \\
12 & Anode/separator boundary. & \\
23 & Cathode/separator boundary. & \\ \hline
\end{tabular}
\caption{Nomenclature of the DFN model.}
\label{tab:nom}
\end{table}

\begin{table}
\centering 
\renewcommand{\arraystretch}{1.3} 
\begin{tabular}{c|l | l}
\hline 
\multicolumn{3}{c}{Parameters for solid diffusion} \\
\hline
$\tau_{c_s}^k$ & $\frac{D_{s,k}}{R_{s,k}^2}$ & Particle time constant.\\
$\theta_{c_s}^k$ & $\frac{dU_k(\cdot)/d c_s^{\text{surf}}}{L_k A_{\text{cc}}}$ & OCP slope per unit volume.\\
\hline 
\multicolumn{3}{c}{Parameters for bulk electrolyte} \\
\hline
$\tau_{c_{e}}^1$ & $\frac{D_{e,1}}{L_1^2}$ & Electrolyte time constant.\\
$\tau_{c_{e}}^2$ & $\frac{D_{e,2}}{L_2^2}$ & Electrolyte time constant.\\
$\tau_{c_{e}}^3$ & $\frac{D_{e,3}}{L_3^2}$ & Electrolyte time constant.\\
$\pi_1$ & $\frac{a_{\text{s},1} \left(1-t_+\right)}{\varepsilon_1 \mathcal{F}}$ & Reactivity term in anode.\\
$\pi_3$ & $\frac{a_{\text{s},3} \left(1-t_+\right)}{\varepsilon_3 \mathcal{F}}$ & Reactivity term in cathode.\\
$\theta_{c_e}^{1,1}$ & $ \sqrt{\frac{D_{e,2}}{D_{e,1}}}$  &  Normalised diffusivity. \\
$\theta_{c_e}^{1,2}$ &   $ \sqrt{\frac{D_{e,2}}{D_{e,3}}}$ & Normalised diffusivity. \\ 
$\theta_{c_e}^{2,1}$ & $ \frac{A_{\text{cc}}K \sigma_1}{\sigma_1+\kappa_1}$ & \\
$\theta_{c_e}^{2,2}$ & $ \frac{A_{\text{cc}}K \kappa_1}{\sigma_1+\kappa_1}$& \\
$\theta_{c_e}^{3,1}$ & $ \frac{A_{\text{cc}}K \sigma_3}{\sigma_3+\kappa_3}$ & \\
$\theta_{c_e}^{3,2}$ & $ \frac{A_{\text{cc}}K \kappa_3}{\sigma_3+\kappa_3}$ & \\
\hline 
\multicolumn{3}{c}{Parameters for overpotential} \\
\hline
$\tau_{\eta}^k$  &   \small{$L_k \sqrt{\left(\frac{\kappa_k+\sigma_k}{\sigma_k\kappa_k}\right)C^{\text{sp}}_k}$} & Double layer time constant. \\
$\theta_{\eta}^{1,k}$  &  \small{$ \sqrt{\frac{L_k^2(\sigma_k+\kappa_k)\mathcal{F}a_{s,k}i_{0,k}^*}{\sigma_k\kappa_kRT}}$} & Normalised reactivity term. \\
$\theta_{\eta}^{2,k} $ &  $A_{\text{cc}}L_k \left(\frac{\kappa_k/\sigma_k + \sigma_k/\kappa_k}{\kappa_k+\sigma_k}\right)$ & A resistance-like term.
\end{tabular}
\caption{Group of twenty one  model parameters that make the electrochemical model structurally identifiable. The overpotential and particle parameters are defined for both the anode and cathode domains set by the index $k \in \{1, \, 3 \}$. In this notation, for instance, $ \tau_{c_s}^1$ is the solid-state diffusion time constant in the anode. }
\label{tab:par}
\end{table}

With these assumptions, a state-space form for the DFN model can then be generated. For this, dynamics for the overpotential are introduced, described in \cite{ong}, by first assuming that $\partial \tilde{c}_e/\partial x \approx 0 $ at the time scales where these dynamics are important (doing so decouples the dynamics of $\phi_{dl}$ from $c_e$) and by adding a double layer term to the divergence equation \eqref{div_eqn}. The obtained linearised state-space form of \eqref{DFN_eqns_orig} is then (\cite{drum_feedback})
\begin{subequations}\label{ss_dyns}\begin{align}
\frac{\partial \tilde{c}_e(x,t)}{\partial t} &= D_e\frac{\partial^2 \tilde{c}_e(x,t) }{\partial x^2}- \frac{a_\text{s} \left(1-t_+\right)}{\varepsilon F}j^*(t),\label{dyns_ce}
\\
\frac{\partial \tilde{u}_s(x,r,t)}{\partial t} &= D_s \frac{\partial ^2\tilde{u}_s(x,r,t)}{\partial r^2},\label{dyns_part}
\\
C^{\text{sp}} \frac{\partial \tilde{\eta}(x,t)}{\partial t} & = \left(\frac{\kappa\sigma}{\kappa+\sigma}\right) \frac{\partial^2 \tilde{\eta}(x,t)}{\partial x^2}-  \frac{a_si_0^*\mathcal{F}}{RT} \tilde{\eta}(x,t). \label{eta_dyns}
\end{align}\end{subequations}
Each of these dynamical equations will be analysed individually to establish structural identifiability.  The electrochemical states of this system are subject to the boundary conditions
\begin{subequations}\label{bcs_part}\begin{align}
\frac{1}{R_s}\frac{\partial \tilde{u}_s(x,r,t)}{\partial r}\Big|_{r = R_s}-\frac{\tilde{u}_s(x,R_\text{s},t)}{R_s^2} & = \frac{-j^{*}(t)}{D_s}, \\
\tilde{u}_s(x,0,t) = 0, 
\end{align}\end{subequations}
\begin{subequations} \label{bcs_eta}\begin{align}\frac{\partial \tilde{\eta}(x,t)}{\partial x}\Big|_{x \in \{0, L\}}  & = -\frac{i(t)}{\sigma} , \\
\frac{\partial \tilde{\eta}(x,t)}{\partial x}\Big|_{x \in \{L_{12}, L_{23}\}}  & =  \frac{i(t)}{\kappa} ,
\end{align} \end{subequations}
\begin{subequations} \label{bcs_ce}\begin{align}
\frac{\partial \tilde{c}_e(x,t)}{\partial x}\Big|_{x \in \{0, L\}} = 0, \label{bcs_ce_no_flux} \\
D_{e,1}\frac{\partial \tilde{c}_e}{\partial x}\Big|_{an} = D_{e,2}\frac{\partial \tilde{c}_e}{\partial x}\Big|_{sep}, ~ x \ = L_{12}, \label{bcs_ce_sep1}\\
D_{e,3}\frac{\partial \tilde{c}_e}{\partial x}\Big|_{cat} = D_{e,2}\frac{\partial \tilde{c}_e}{\partial x}\Big|_{sep}, ~ x \ = L_{23}, \label{bcs_ce_sep2} \\
\tilde{c}_e(x,t) \text{ is continuous on } x \in \{L_{12}, L_{23}\}. \label{bcs_ce_cont}
\end{align}\end{subequations}
Here, the subscript $an$ denotes a derivative normal to the boundary of the anode, $sep$ is normal to the separator and $cat$ is normal to the cathode. 

\subsection{Voltage expression}
The voltage is defined as the difference in solid-phase potentials between each current collector but requires some reformulation to be expressed in terms of the electrochemical states. For this, consider
\begin{subequations}\begin{align}
v(t)  & = \phi_s(L,t)-\phi_s(0,t) - R_\text{ctc} i(t) \\
& = {\phi}_{dl}(L,t) - {\phi}_{dl}(0,t) + \phi_e(L,t) - \phi_e(0,t) - R_\text{ctc} i(t)\\
&= {\phi}_{dl}(L,t) - {\phi}_{dl}(0,t) + \int_0^{L} \frac{\partial \phi_e(x,t)}{\partial x} ~dx - R_\text{ctc} i(t)
\end{align}
with $ \frac{\partial \phi_e}{\partial x}$ evaluated from MacInnes' equation \eqref{ie1} 
\begin{align}\label{volt_int}
v(t)& = {\phi}_{dl}(L,t) - {\phi}_{dl}(0,t) \\
& + \int_0^{L} \frac{-i_e(x,t)}{\kappa} + K\frac{\partial \tilde{c}_e(x,t)}{\partial x}~dx- R_\text{ctc} i(t). \nonumber
\end{align}
\end{subequations}

Analysing the integral in \eqref{volt_int} piece-wise across each of the three domains $\Omega_1,$ $\Omega_2$, $\Omega_3$ using the relationship
\begin{subequations}\begin{align}
i_e(x,t)  & = \frac{\sigma_k \kappa_k}{\sigma_k+\kappa_k} \frac{\partial \phi_{dl} }{\partial x}   + \frac{K\sigma_k \kappa_k}{\sigma_k+\kappa_k} \frac{\partial \tilde{c}_e }{\partial x}  + \frac{ \kappa_ki(t)}{\sigma_k+\kappa_k}  ,\nonumber \\ &  \quad \quad  \quad x \in \Omega_k, k \in \{1,3\}, 
\\
i_e(x,t)  & = i(t), ~ x \in \Omega_2,
\end{align}\end{subequations}
 allows the voltage to be expressed in terms of the electrochemical states
 \begin{align}\label{v_phi}
v(t)& = {\phi}_{dl}(L,t) - {\phi}_{dl}(0,t) \\
& + \int_{\Omega_1} -\frac{\sigma_1 }{\sigma_1+\kappa_1} \frac{\partial \phi_{dl} }{\partial x}   + \frac{K\kappa_1 }{\sigma_1+\kappa_1} \frac{\partial \tilde{c}_e }{\partial x}  - \frac{ i(t)}{\sigma_1+\kappa_1} ~dx \nonumber
\\
& + \int_{\Omega_3} -\frac{\sigma_3 }{\sigma_k+\kappa_3} \frac{\partial \phi_{dl} }{\partial x}   + \frac{K\kappa_3 }{\sigma_3+\kappa_3} \frac{\partial \tilde{c}_e }{\partial x}  - \frac{ i(t)}{\sigma_3+\kappa_3} ~dx \nonumber
\\
& + \int_{\Omega_2}  K \frac{\partial \tilde{c}_e }{\partial x}  - \frac{ i(t)}{\kappa_k} ~dx
- R_\text{ctc} i. \nonumber
\end{align}
 However, in this paper, the analysis is concerned with local perturbations of the voltage around the equilibrium of the open circuit voltage (OCV). These local perturbations are found by substituting the expression for $\phi_{dl}$ from the overpotential definition \eqref{eta} into \eqref{v_phi} and using the approximation
\begin{align}
U(c_s^{\text{surf}}(x,t)) \approx U(c_s^{\text{surf},*})  + \frac{1}{R_s}\frac{dU\left(c_s^{\text{surf}}\right)}{dc_s^{\text{surf}}}\Big|_{c_s^{\text{surf}} = c_s^{\text{surf},*}} \tilde{u}_s^{\text{surf}}.
\end{align}
 The voltage can then be expressed as 
\begin{align}
v(t) = OCV + \tilde{v}(t),
\end{align}
with the OCV defined from the $U(c_s^{\text{surf},*})$ terms and the perturbations being
\begin{subequations}\begin{align}
\tilde{v}(t) = v_{\eta}(t) +v_{u_s^{\text{surf}}}(t) + v_{c_e}(t) - R_\text{res} i(t)
\end{align}
where 
\begin{align} \label{eqns:voltage}
v_{\eta}(t)   &   = - \frac{\kappa_1}{\kappa_1+\sigma_1}\tilde{\eta}(0,t)  
 - \frac{\sigma_1}{\kappa_1+\sigma_1}\tilde{\eta}(L_{12},t) \
 \\ &  \nonumber
 +\frac{\sigma_3}{\kappa_3+\sigma_3} \tilde{\eta}(L_{23},t)
 +\frac{\kappa_3}{\kappa_3+\sigma_3} \tilde{\eta}(L,t),
\\
 v_{u_s^{\text{surf}}}(t)  & =  - \frac{\frac{\kappa_1}{R_s}\frac{dU_1}{d c_s^{\text{surf}}}}{\kappa_1+\sigma_1} \tilde{u}_s^{\text{surf}}(0,t) 
 - \frac{\frac{\sigma_1}{R_s}\frac{dU_1}{d c_s^{\text{surf}}}}{\kappa_1+\sigma_1} \tilde{u}_s^{\text{surf}}(L_{12},t)  \nonumber
 \\ &  
 +\frac{\frac{\sigma_3}{R_s}\frac{dU_3}{d c_s^{\text{surf}}}}{\kappa_3+\sigma_3}  \tilde{u}_s^{\text{surf}}(L_{23},t)
 +\frac{\frac{\kappa_3}{R_s}\frac{dU_3}{d c_s^{\text{surf}}}}{\kappa_3+\sigma_3}  \tilde{u}_s^{\text{surf}}(L,t),
\\
v_{ce}(t) &  = -\frac{K\kappa_1 }{\sigma_1+\kappa_1} \tilde{c}_e(0,t) -\frac{K\sigma_1 }{\sigma_1+\kappa_1}\tilde{c}_e(L_{12},t) \nonumber
\\ &
+\frac{K\sigma_3 }{\sigma_3+\kappa_3}\tilde{c}_e(L_{23},t) + \frac{K\kappa_3 }{\sigma_3+\kappa_3}\tilde{c}_e(L,t)\label{v_ce} ,
\\
 R_\text{res} & =  R_\text{ctc}    + \frac{ L_1}{\sigma_3+\kappa_3} + \frac{ L_2}{\kappa_2} + \frac{ L_3}{\sigma_3+\kappa_3}.
\end{align}\end{subequations}

Under the constant reactivity assumption of the SPM \eqref{j_const}, which holds for cells with sufficiently high electronic conductivities, then the voltage from the particles' surfaces can be simplified to
\begin{align} \label{v_part2}
 v_{u_s^{\text{surf}}}  \approx \frac{1}{R_{s,3}} \frac{dU_3}{d c_s^{\text{surf}}} \tilde{u}_s^{\text{surf}}(L_{23},t)  - \frac{1}{R_{s,1}} \frac{dU_1}{d c_s^{\text{surf}}} \tilde{u}_s^{\text{surf}}(L_{12},t),
\end{align}
allowing the particle dynamics to be described by a single particle.

\subsection{Structural Identifiability}

This paper is concerned with structural identifiability of this system (\cite{structural}). Structural identifiability is defined as (\cite{ljung,randles}):
\begin{definition}[Structural identifiability]
 Consider a model structure $\mathcal{M}$ with the transfer function $H(s, \theta)$ parametrised by $\theta \in D \subset \mathbb{R}^n$ where $n$ denotes
the number of parameters of the model. The identifiability
equation for $\mathcal{M}$ is given by:
\begin{align} \label{eq:struct}
H(s, \theta) = H(s, \theta^*)~ \text{for almost all }s
\end{align}
where $\theta, \theta^* \in \mathcal{D}$. The model structure $\mathcal{M}$ is said to be
\begin{itemize}
\item globally identifiable if \eqref{eq:struct} has a unique solution in $\mathcal{D}$,
\item locally identifiable if \eqref{eq:struct} has a finite number of solutions in $\mathcal{D}$,
\item unidentifiable if \eqref{eq:struct} has a infinite number of solutions
in $\mathcal{D}$.
\end{itemize}

\end{definition}
If the model is globally identifiable then it is said to be structurally identifiable and the parameter estimation problem is well-posed (meaning that it admits a unique solution). To verify structural identifiability of the considered model, the minimum number of parameters needed to uniquely characterise the impedance functions of each electrochemical state is determined.

\section{Impedance analysis for structural identifiability} 
The analysis of the previous section, under the given assumptions, decomposed the DFN model into three components with three distinct timescales: solid-state diffusion of $u_s$ (slow); bulk electrolyte diffusion (mid) and relaxation of overpotentials (fast).  Structural identifiability of each of the three decoupled timescales will now be analysed individually.

\subsection{Low-frequency range: Solid-state diffusion}
Solid-state diffusion within the active material particles dominates the low frequency response (when $\omega \ll 1$ rad s$^{-1}$), with the structural identifiability of these dynamics detailed in \cite{adrien_ident} and repeated here for completeness. These dynamics are defined by the radial diffusion of \eqref{dyns_part}, \eqref{bcs_part}, \eqref{v_part2} having Laplace transform
\begin{align}
\frac{\bar{\tilde{c}}^{surf}_{s}(x,s)}{\bar{j}^*(s)} = \left(\frac{R_s ^2}{a_s\mathcal{F}D_s}\right)\left( \frac{\tanh\left( \sqrt{\frac{s}{\tau_{c_s}}}\right)}{\tanh\left(\sqrt{\frac{s}{\tau_{c_s}}}\right)- \sqrt{\frac{s}{\tau_{c_s}}}}\right). \label{SPMgain}
\end{align}
Applying the constant reactivity approximation of \eqref{j_const} to \eqref{SPMgain} gives the transfer function of the SPM
\begin{align}
\frac{ \bar{v}_{u_s^{\text{surf}}}}{\bar{I}(s)}  & = \theta_{c_s,3} \tau_{c_s,3}\left( \frac{\tanh\left( \sqrt{\frac{s}{\tau_{c_s,3}}}\right)}{\tanh\left(\sqrt{\frac{s}{\tau_{c_s,3}}}\right)- \sqrt{\frac{s}{\tau_{c_s,3}}}}\right) \\ & 
-\theta_{c_s,1} \tau_{c_s,1}\left( \frac{\tanh\left( \sqrt{\frac{s}{\tau_{c_s,1}}}\right)}{\tanh\left(\sqrt{\frac{s}{\tau_{c_s,1}}}\right)- \sqrt{\frac{s}{\tau_{c_s,1}}}}\right)
\end{align}
with $\theta_{c_s^{\text{surf}}}^k =  \frac{dU_k/d c_s^{\text{surf}}}{L_k A_{cc}}$ and $\tau_{c_s}^k = R_{s,k}/\sqrt{D_{s,k}}$ where $k \in \{1, 3 \}$.
These are the four parameters that uniquely map the input current to variations in particle surface concentrations, under the above assumptions.  

\subsection{Mid-frequency range: Movement of bulk electrolyte}
Moving up the frequency range to $\omega \approx 1$ rad s$^{-1}$, polarisation of the bulk electrolyte becomes the dominant electrochemical effect. Structural identifiability of the electrolyte requires a transfer function for \eqref{dyns_ce} and \eqref{v_ce}. Unfortunately, due to the boundary conditions \eqref{bcs_ce_no_flux}-\eqref{bcs_ce_cont} coupling the electrolyte's solutions across various domains, the impedance analysis of this state is more complex to analyse than the other effects considered in the model. To alleviate some of this complexity, the uniform reactivity approximation \eqref{j_lin} is applied to \eqref{dyns_ce}, converting the dynamics in the electrodes to 
\begin{subequations}\label{c_e_dyns} \begin{align}
\frac{\partial \tilde{c}_e(x,t)}{\partial t}  & = D_{e,1}  \frac{\partial^2 \tilde{c}_e(x,t)}{\partial x^2} + \pi_1  i(t), \forall x \in \Omega_1, \\
\frac{\partial \tilde{c}_e(x,t)}{\partial t}  & = D_{e,2}  \frac{\partial^2 \tilde{c}_e(x,t)}{\partial x^2}, \forall x \in \Omega_2, \\
\frac{\partial \tilde{c}_e(x,t)}{\partial t}  & = D_{e,3}  \frac{\partial^2 \tilde{c}_e(x,t)}{\partial x^2} -\pi_3  i(t), \forall x \in \Omega_3,
\end{align}\end{subequations}
where $\pi_k = \frac{a_{\text{s},k} \left(1-t_+\right)}{\varepsilon_k \mathcal{F}}$, $k  \in \{1,3\}$ is a parameter for the reactivity. 

The co-ordinate transformation $\gamma$ is introduced, being defined in each electrode as 
\begin{subequations} \begin{align}
\gamma(x,t) = \tilde{c}_e(x,t) - \pi_1\int^t_0 i(\tau) \, d \tau, ~ x \in \Omega_1, \\
\gamma(x,t) = \tilde{c}_e(x,t) + \pi_3\int^t_0 i(\tau) \, d \tau, ~ x \in \Omega_3.
\end{align}  \end{subequations}
Then, in the electrodes,
\begin{subequations}\begin{align}
\frac{\partial \gamma(x,t)}{\partial t} & = \frac{\partial \tilde{c}_e(x,t)}{\partial t} + \pi_1 i(t), \forall x \in \Omega_1, \\
\frac{\partial \gamma(x,t)}{\partial t} & = \frac{\partial \tilde{c}_e(x,t)}{\partial t} - \pi_3 i(t),~ \forall x \in \Omega_3, \\
\frac{\partial^2 \gamma(x,t)}{\partial x^2}  & = \frac{\partial^2 \tilde{c}_e(x,t)}{\partial x^2},
\end{align} \end{subequations}
with the dynamics in this co-ordinate system then converting \eqref{c_e_dyns} into the simpler, unforced diffusion equation
\begin{align}\label{gamma_dyns}
\frac{\partial \gamma(x,t)}{\partial t} = D_{e,k}  \frac{\partial^2 \gamma(x,t)}{\partial x^2}, \, x \in \{\Omega_k\}, \, k \in \{1, \, 3\},
\end{align}
where the forcing reactivity term in \eqref{c_e_dyns} has been moved into the boundary conditions via \eqref{bcs_ce_cont}. 

In each domain, the Laplace transform solution for \eqref{gamma_dyns} (assuming zero initial conditions) is
\begin{subequations}\label{gamma_s}\begin{align}
\bar{\gamma}(x,s)  & = A_1(s)e^{\frac{\sqrt{s}x}{\sqrt{D_{e,1}}}} + B_1(s)e^{\frac{-\sqrt{s}x}{\sqrt{D_{e,1}}}}, x \in \Omega_1, \\
\bar{\tilde{c}}_{e}(x,s)  & = A_2(s)e^{\frac{\sqrt{s}\left(x-L_{12}\right)}{\sqrt{D_{e,2}}}} + B_2(s)e^{\frac{-\sqrt{s}\left(x-L_{12}\right)}{\sqrt{D_{e,2}}}} , x \in \Omega_2, \\
\bar{\gamma}(x,s)  & = A_3(s)e^{\frac{\sqrt{s}\left(x-L_{23}\right)}{\sqrt{D_{e,3}}}} + B_3(s)e^{\frac{-\sqrt{s}\left(x-L_{23}\right)}{\sqrt{D_{e,3}}}}, x \in \Omega_3, 
\end{align}\end{subequations}
with spatial derivatives
\small{\begin{subequations}\begin{align}
\frac{\partial \bar{\gamma}}{\partial x}  & = \frac{A_1(s)\sqrt{s}}{\sqrt{D_{e,1}}}e^{\frac{\sqrt{s}x}{\sqrt{D_{e,1}}}} - \frac{B_1(s)\sqrt{s}}{\sqrt{D_{e,1}}}e^{\frac{-\sqrt{s}x}{\sqrt{D_{e,1}}}}, x \in \Omega_1,  \\
\frac{\partial \bar{\tilde{c}}_{e}}{\partial x}  & = \frac{A_2(s)\sqrt{s}}{\sqrt{D_{e,2}}}e^{\frac{\sqrt{s}\left(x-L_{12}\right)}{\sqrt{D_{e,2}}}} - \frac{B_2(s)\sqrt{s}}{\sqrt{D_{e,2}}}e^{\frac{-\sqrt{s}\left(x-L_{12}\right)}{\sqrt{D_{e,2}}}}, x \in \Omega_2,  \\
\frac{\partial \bar{\gamma}}{\partial x}  & = \frac{A_3(s)\sqrt{s}}{\sqrt{D_{e,3}}}e^{\frac{\sqrt{s}\left(x-L_{23}\right)}{\sqrt{D_{e,3}}}} - \frac{B_3(s)\sqrt{s}}{\sqrt{D_{e,3}}}e^{\frac{-\sqrt{s}\left(x-L_{23}\right)}{\sqrt{D_{e,3}}}}, x \in \Omega_3 .
\end{align}\end{subequations}}\normalsize
 The six boundary conditions for the electrolyte \eqref{bcs_ce} allow the six solution coefficients $(A_1, \,B_1,\, A_2, \, B_2, \, A_3, \, B_3)$ of \eqref{gamma_s} to be computed.

Beginning with the no-flux condition at the current collectors \eqref{bcs_ce_no_flux}, then
\begin{subequations}\label{Bs}\begin{align}
 A_1(s)  & = B_1(s), \label{B1} \\
 B_3(s)  & = A_3(s)e^{\frac{2\sqrt{s}L_3}{\sqrt{D_{e,3}}}}. \label{B2}
\end{align}\end{subequations}
The constant flux conditions \eqref{bcs_ce_sep1}, \eqref{bcs_ce_sep2} at the electrode/separator means
\begin{subequations}\label{A1_A3}\small{\begin{align}
  & A_1(s)= \sqrt{\frac{D_{e,2}}{D_{e,1}}}\left(\frac{A_2(s) - B_2(s)}{e^{\sqrt{\frac{s}{D_{e,1}}}L_1} -e^{-\sqrt{\frac{s}{D_{e,1}}}L_1}}\right), \label{A1}
 \\
 &  A_3(s) =  \sqrt{\frac{D_{e,2}}{D_{e,3}}}\left(\frac{A_2(s)e^{\sqrt{\frac{s}{D_{e,2}}}L_2} -B_2(s)e^{-\sqrt{\frac{s}{D_{e,2}}}L_2}}{1 - e^{\frac{2\sqrt{s}L_3}{\sqrt{D_{e,3}}}}}\right).\label{A3}
\end{align}}\end{subequations}
Lastly, solution continuity at the separator interface \eqref{bcs_ce_cont} ensures
\begin{subequations}\begin{align} 
\pi_1\frac{\bar{i}(s)}{s} + &  A_1(s)\left(e^{\frac{\sqrt{s}L_1}{\sqrt{D_{e,1}}}} + e^{-\frac{\sqrt{s}L_1}{\sqrt{D_{e,1}}}}\right)  = 
A_2(s) + B_2(s), \label{cont1}
\end{align}
\begin{align}
A_2(s)e^{\frac{\sqrt{s}L_2}{\sqrt{D_{e,2}}}} & + B_2(s)e^{-\frac{\sqrt{s}L_2}{\sqrt{D_{e,2}}}} \label{cont2}
\\ & 
= A_3(s) \left(1+e^{\frac{2\sqrt{s}L_3}{\sqrt{D_{e,3}}}}\right)-\pi_3\frac{\bar{i}(s)}{s}. \nonumber
\end{align}\end{subequations}
We then have four equations to solve for the four remaining unknowns $A_1(s),\,A_3(s),\, A_2(s)$ and $ B_2(s)$.
 
Substituting \eqref{A1} into the continuity equation \eqref{cont1} gives
\begin{subequations}\begin{align} \label{k_Is1}
\pi_1\frac{\bar{i}(s)}{s} = (1-p_1(s))A_2(s) + (1+p_1(s))B_2(s)
\end{align}
with 
\begin{align}
p_1(s) = \sqrt{\frac{D_{e,2}}{D_{e,1}}}\frac{\left(e^{\sqrt{\frac{s}{D_{e,1}}}L_1} + e^{-\sqrt{\frac{s}{D_{e,1}}}L_1}\right)}{\left(e^{\sqrt{\frac{s}{D_{e,1}}}L_1} -e^{-\sqrt{\frac{s}{D_{e,1}}}L_1}\right)}.
\end{align}\end{subequations}
The variables $p_k(s),$  $k = 1, \, \dots \,, \, 7$ are introduced to simplify the algebra. Similarly, with \eqref{A3}, the continuity condition \eqref{cont2} at the right hand side can be written
\begin{subequations}\begin{align} \label{k_Is2}
-\pi_3\frac{\bar{i}(s)}{s} = p_2(s)A_2(s) + p_2(s)e^{\frac{-2\sqrt{s}L_2}{\sqrt{D_{e,2}}}}B_2(s)
\end{align}
where
\begin{align}
p_2(s)  & = e^{\frac{\sqrt{s}L_2}{\sqrt{D_{e,2}}}} \left(1 -\sqrt{\frac{D_{e,2}}{D_{e,3}}}\left(\frac{1+e^{\frac{2\sqrt{s}L_3}{\sqrt{D_{e,3}}}}}{1 - e^{\frac{2\sqrt{s}L_3}{\sqrt{D_{e,3}}}}}\right)\right),
  \\
p_3(s)  & = e^{-\frac{\sqrt{s}L_2}{\sqrt{D_{e,2}}}} \left(1  +\sqrt{\frac{D_{e,2}}{D_{e,3}}}\left(\frac{1+e^{\frac{2\sqrt{s}L_3}{\sqrt{D_{e,3}}}}}{1 - e^{\frac{2\sqrt{s}L_3}{\sqrt{D_{e,3}}}}}\right)\right).
\end{align}\end{subequations}
Equating \eqref{k_Is1} and \eqref{k_Is2} gives an expression for $A_2(s)$
\begin{subequations}\begin{align}
A_2(s) &  = \left( \frac{\pi_1  p_3(s) + \pi_3(1+p_1(s))}{-\pi_1p_2(s)-\pi_3(1-p_1(s))} \right)B_2(s).
\\ & = p_4(s) B_2(s).
\end{align}\end{subequations}
Subbing back into either \eqref{k_Is1} or \eqref{k_Is2} gives $B_2(s)$ in terms of $\bar{i}(s)$
\begin{subequations}\begin{align}
 B_2(s)  & = \frac{\pi_1}{s( p_2(s)(1-p_1(s))+1+p_1(s))}\bar{i}(s)= p_5(s)\bar{i}(s), \\
  & = \frac{-\pi_3}{s \left(p_2(s)p_3(s) + p_3(s)\right)}\bar{i}(s),
\end{align}\end{subequations}
and so
\begin{align}
 A_2(s) =   p_4(s)p_5(s)\bar{i}(s).
\end{align}
The remaining unknowns $A_1(s)$ and $A_3(s)$ can then be computed from \eqref{A1_A3}
\begin{subequations}\label{A1_A32}\small{\begin{align}
  & A_1(s)= \sqrt{\frac{D_{e,2}}{D_{e,1}}}\left(\frac{p_4(s)(p_5(s)-1)}{e^{\sqrt{\frac{s}{D_{e,1}}}L_1} -e^{-\sqrt{\frac{s}{D_{e,1}}}L_1}}\right)\bar{i}(s), \label{A1new}
 \\
 &  A_3(s) =  \sqrt{\frac{D_{e,2}}{D_{e,3}}}\left(\frac{p_4(s)\left(p_5(s)e^{\frac{\sqrt{s}L_2}{\sqrt{D_{e,2}}}} -e^{-\frac{\sqrt{s}L_2}{\sqrt{D_{e,2}}}}\right)}{1 - e^{2\sqrt{\frac{{s}}{{D_{e,3}}}}L_3}}\right)\bar{i}(s),\label{A3new}
\end{align}}\end{subequations}
or, more compactly,
\begin{subequations}\label{A1_A32}\small{\begin{align}
  & A_1(s)= p_6(s)\bar{i}(s), \label{A1new}
 \\
 &  A_3(s) = p_7(s)\bar{i}(s).\label{A3new}
\end{align}}\end{subequations}
The impedance function of the bulk electrolyte can then be expressed by
\begin{align}
\frac{\bar{v}_{c_e}(s)}{\bar{I}(s)} &  = -\frac{Kp_6(s) }{\sigma_1+\kappa_1}\left(2\kappa_1 +\sigma_1 \left(e^{\frac{\sqrt{s}L_1}{\sqrt{D_{e,1}}}} +e^{\frac{-\sqrt{s}L_1}{\sqrt{D_{e,1}}}} \right)\right)\nonumber
\\ &
+\frac{K p_7(s)}{\sigma_3+\kappa_3}\left(\sigma_3 + \sigma_3e^{\frac{2\sqrt{s}L_3}{\sqrt{D_{e,3}}}} + 2\kappa_3 e^{\frac{\sqrt{s}L_3}{\sqrt{D_{e,3}}}} \right) \nonumber
\\ &
 + K\frac{(\pi_3+\pi_1)}{s} ,
\end{align}
which is uniquely identified by the eleven parameters in Table \ref{tab:par}. These parameters relate to the the ionic time constants, the reactivity terms and the relative diffusivities of each adjacent domain. 

\subsection{High-frequency response: Overpotential dynamics}
The high-frequency response of the cell follows from the localised overpotential dynamics \eqref{eta_dyns} coupled to the boundary conditions \eqref{bcs_eta}. In $x \in \Omega_k$, $k \in \{1,3\}$, the Laplace transform for this system was given  in \cite{ong} as
\begin{align} \label{eta_s}
 & \frac{\bar{\tilde{\eta}}(x,s)}{\bar{i}(s)} = \frac{L_k}{\upsilon_k(s) \sinh(\upsilon_k(s))} 
 \\ & \times \left[ \frac{1}{\sigma_k}\cosh\left(\frac{\upsilon_k(s)x}{L_k}\right)
+\frac{1}{\kappa_k}\cosh\left(\upsilon_k(s)\left(1-\frac{x}{L_k}\right)\right) \right], \nonumber
\end{align}
with $\upsilon_k(s)$ given by
\begin{align}
\frac{\upsilon_k^2}{L_k^2} = \left(\frac{1}{\kappa_k}+\frac{1}{\sigma_k}\right)\left( \frac{\mathcal{F}a_{s,k}i_{0,k}^*}{RT}+C^{\text{sp}}_ks\right).
\end{align}
From \eqref{eta_s}, the transfer function of the overpotential dynamics, corresponding to the charge transfer resistance, can be computed as

\begin{align}
\frac{v_{\eta}(s)}{\bar{i}(s)} = &  \left(\frac{\kappa_3/\sigma_3 + \sigma_3/\kappa_3}{\kappa_3+\sigma_3}\right)\frac{L_3 \coth(v_3(s))}{ v_3(s) }  \nonumber
\\
 & - \left(\frac{\kappa_1/\sigma_1 + \sigma_1/\kappa_1}{\kappa_1+\sigma_1}\right)\frac{L_1 \coth(v_1(s)}{ v_1(s) }  .
\end{align}
This function is parameterised by 
\begin{subequations}\begin{align}
\theta_{\eta}^{1,k}  & = L_k \sqrt{\left(\frac{1}{\kappa_k}+\frac{1}{\sigma_k}\right)\frac{\mathcal{F}a_{s,k}i_{0,k}^*}{RT}},~k \in \{1, 3\}, \\
\theta_{\eta}^{2,k} & = A_{\text{cc}}L_k \left(\frac{\kappa_k/\sigma_k + \sigma_k/\kappa_k}{\kappa_k+\sigma_k}\right), ~k \in \{1, 3\}, \\ 
\tau_{\eta}^k    & =  L_k^2 {\left(\frac{1}{\kappa_k}+\frac{1}{\sigma_k}\right)C^{\text{sp}}_k}, ~k \in \{1, 3\},
\end{align} \end{subequations}
so that
\begin{align}
\upsilon_k = \theta^{\eta}_{1,k} + \sqrt{\tau_{\eta,k}}  \sqrt{s}
\end{align}
and
\begin{align}
\frac{v_{\eta}(s)}{\bar{I}(s)} = &  \theta^{\eta}_{2,3}\frac{ \coth(v_3(s))}{ v_3(s) } - \theta^{\eta}_{2,1}\frac{ \coth(v_1(s))}{ v_1(s) }  .
\end{align}
Six parameters then uniquely determine the response due to heterogeneous electrode overpotentials.  

\subsection{Combined analysis}
When combined, these results show that the current to voltage mapping of the considered decoupled and linearised battery model is characterised uniquely by the 21 parameters of Table \ref{tab:par}. The model is then structurally identifiable with respect to these parameters and a unique solution to the parameter estimation problem for this model then exists when current-voltage data is used. The obvious next step of this research is then to develop algorithms to estimate values for these identifiable parameters from data, with the key message of this paper being that the feasibility of obtaining unique parameter estimates for these electrochemical models should be considered before any algorithms are applied. 

\section*{Conclusions}
A structural identifiability analysis of a decoupled and linearised Doyle-Fuller-Newman Li-ion battery electrochemical model was applied. It was shown that the model is structurally identifiable from a group of 21 parameters (composed of electrochemical quantities like the conductivities and lengths), with these parameters uniquely characterising the impedance function of this model. The parameter estimation problem for this model is then well-posed with respect to these groups of parameters. Future work will aim to exploit this result to develop an algorithm to recursively estimate the parameter values for pseudo-2D battery models from generic data.

\bibliography{bibliog}

\begin{thebibliography}{20}
\providecommand{\natexlab}[1]{#1}
\providecommand{\url}[1]{\texttt{#1}}
\providecommand{\urlprefix}{URL }
\expandafter\ifx\csname urlstyle\endcsname\relax
  \providecommand{\doi}[1]{doi:\discretionary{}{}{}#1}\else
  \providecommand{\doi}{doi:\discretionary{}{}{}\begingroup
  \urlstyle{rm}\Url}\fi

\bibitem[{Alavi et~al.(2016)Alavi, Mahdi, Payne, and Howey}]{randles}
Alavi, S.M.M., Mahdi, A., Payne, S.J., and Howey, D.A. (2016).
\newblock Identifiability of generalized {R}andles circuit models.
\newblock \emph{IEEE Transactions on Control Systems Technology}, 25(6),
  2112--2120.

\bibitem[{Atlung et~al.(1979)Atlung, West, and Jacobsen}]{SPM2}
Atlung, S., West, K., and Jacobsen, T. (1979).
\newblock Dynamic aspects of solid solution cathodes for electrochemical power
  sources.
\newblock \emph{Journal of The Electrochemical Society}, 126(8), 1311--1321.

\bibitem[{Bellman and {\AA}str{\"o}m(1970)}]{structural}
Bellman, R. and {\AA}str{\"o}m, K.J. (1970).
\newblock On structural identifiability.
\newblock \emph{Mathematical biosciences}, 7(3-4), 329--339.

\bibitem[{Bizeray et~al.(2018)Bizeray, Kim, Duncan, and Howey}]{adrien_ident}
Bizeray, A.M., Kim, J.H., Duncan, S.R., and Howey, D.A. (2018).
\newblock Identifiability and parameter estimation of the single particle
  lithium-ion battery model.
\newblock \emph{IEEE Transactions on Control Systems Technology}, (99), 1--16.

\bibitem[{Chaturvedi et~al.(2010)Chaturvedi, Klein, Christensen, Ahmed, and
  Kojic}]{chaturvedi}
Chaturvedi, N.A., Klein, R., Christensen, J., Ahmed, J., and Kojic, A. (2010).
\newblock Algorithms for advanced battery-management systems.
\newblock \emph{IEEE Control Systems}, 30(3), 49--68.

\bibitem[{Curry(2017)}]{curry}
Curry, C. (2017).
\newblock Lithium-ion battery costs and market.
\newblock \emph{Bloomberg New Energy Finance}, 5.

\bibitem[{Doyle et~al.(1993)Doyle, Fuller, and Newman}]{Newman1}
Doyle, M., Fuller, T.F., and Newman, J. (1993).
\newblock Modeling of galvanostatic charge and discharge of the
  lithium/polymer/insertion cell.
\newblock \emph{Journal of the Electrochemical Society}, 140(6), 1526--1533.

\bibitem[{Drummond et~al.(2019)Drummond, Bizeray, Howey, and
  Duncan}]{drum_feedback}
Drummond, R., Bizeray, A.M., Howey, D.A., and Duncan, S.R. (2019).
\newblock A feedback interpretation of the {D}oyle-{F}uller-{N}ewman
  lithium-ion battery model.
\newblock \emph{IEEE Transactions on Control Systems Technology}.

\bibitem[{Forman et~al.(2012)Forman, Moura, Stein, and Fathy}]{moura_genetic}
Forman, J.C., Moura, S.J., Stein, J.L., and Fathy, H.K. (2012).
\newblock Genetic identification and {F}isher identifiability analysis of the
  {D}oyle-{F}uller-{N}ewman model from experimental cycling of a {LiFePO4}
  cell.
\newblock \emph{Journal of Power Sources}, 210, 263--275.

\bibitem[{Kang and Ceder(2009)}]{fast}
Kang, B. and Ceder, G. (2009).
\newblock Battery materials for ultrafast charging and discharging.
\newblock \emph{Nature}, 458(7235), 190.

\bibitem[{Lai et~al.(2019)Lai, Jangra, Ahn, Kim, Joe, and Lin}]{xinfan}
Lai, Q., Jangra, S., Ahn, H.J., Kim, G., Joe, W.T., and Lin, X. (2019).
\newblock Analytical sensitivity analysis for battery electrochemical
  parameters.
\newblock In \emph{American Control Conference, Philadelphia, PA}, 890--896.
  IEEE.

\bibitem[{Ljung(2001)}]{ljung}
Ljung, L. (2001).
\newblock System identification.
\newblock \emph{Wiley Encyclopedia of Electrical and Electronics Engineering}.

\bibitem[{Marquis et~al.(2019)Marquis, Sulzer, Timms, Please, and
  Chapman}]{marquis}
Marquis, S.G., Sulzer, V., Timms, R., Please, C.P., and Chapman, S.J. (2019).
\newblock An asymptotic derivation of a single particle model with electrolyte.
\newblock \emph{arXiv preprint arXiv:1905.12553}.

\bibitem[{Moura et~al.(2017)Moura, Argomedo, Klein, Mirtabatabaei, and
  Krstic}]{SPMe}
Moura, S.J., Argomedo, F.B., Klein, R., Mirtabatabaei, A., and Krstic, M.
  (2017).
\newblock Battery state estimation for a single particle model with electrolyte
  dynamics.
\newblock \emph{IEEE Transactions on Control Systems Technology}, 25(2),
  453--468.

\bibitem[{Ning and Popov(2004)}]{SPM1}
Ning, G. and Popov, B.N. (2004).
\newblock Cycle life modeling of lithium-ion batteries.
\newblock \emph{Journal of The Electrochemical Society}, 151(10), A1584--A1591.

\bibitem[{Ong and Newman(1999)}]{ong}
Ong, I.J. and Newman, J. (1999).
\newblock Double-layer capacitance in a dual lithium ion insertion cell.
\newblock \emph{Journal of The Electrochemical Society}, 146(12), 4360--4365.

\bibitem[{Park et~al.(2018)Park, Kato, Gima, Klein, and Moura}]{park}
Park, S., Kato, D., Gima, Z., Klein, R., and Moura, S. (2018).
\newblock Optimal experimental design for parameterization of an
  electrochemical lithium-ion battery model.
\newblock \emph{Journal of The Electrochemical Society}, 165(7), A1309--A1323.

\bibitem[{Pozzi et~al.(2018)Pozzi, Ciaramella, Volkwein, and Raimondo}]{pozzi}
Pozzi, A., Ciaramella, G., Volkwein, S., and Raimondo, D.M. (2018).
\newblock Optimal design of experiments for a lithium-ion cell: {P}arameters
  identification of an isothermal single particle model with electrolyte
  dynamics.
\newblock \emph{Industrial \& Engineering Chemistry Research}, 58(3),
  1286--1299.

\bibitem[{Richardson et~al.(2019)Richardson, Korotkin, Castle, and
  Foster}]{giles}
Richardson, G., Korotkin, I., Castle, R., and Foster, J. (2019).
\newblock Generalised single particle models for high-rate operation of graded
  lithium-ion electrodes: systematic derivation and validation.
\newblock \emph{arXiv preprint arXiv:1907.09410}.

\bibitem[{Sikha and White(2007)}]{sikha}
Sikha, G. and White, R.E. (2007).
\newblock Analytical expression for the impedance response of an insertion
  electrode cell.
\newblock \emph{Journal of The Electrochemical Society}, 154(1), A43--A54.

\end{thebibliography}


\end{document}